**Article type: Original Paper**

**Experimental demonstration of 𝒫𝒯-symmetric stripe lasers**

*Zhiyuan Gu[1, †], Nan Zhang[1, †], Quan Lyu[1, †], Meng Li[1], Ke Xu[1], Shumin Xiao[2], Qinghai Song[1, 3, *]*

*Corresponding Author: Qinghai.song@hitsz.edu.cn

[†]*These authors contribute equally to this work.*

1. Integrated Nanoscience Lab, Department of Electrical and Information Engineering, Harbin Institute of Technology, Shenzhen, 518055, China
2. Integrated Nanoscience Lab, Department of Material Science and Engineering, Harbin Institute of Technology, Shenzhen, 518055, China
3. State Key Laboratory on Tunable laser Technology, Harbin Institute of Technology, Harbin, 158001, China

**Abstract:**

Recently, the coexistence of parity-time (PT) symmetric laser and absorber has gained tremendous research attention. While the PT symmetric absorber has been observed in microwave metamaterials, the experimental demonstration of PT symmetric laser is still absent. Here we experimentally study PT-symmetric laser absorber in stripe waveguide. Using the concept of PT symmetry to exploit the light amplification and absorption, PT-symmetric laser absorbers have been successfully obtained. Different from the single-mode PT symmetric lasers, the PT-symmetric stripe lasers have been experimentally confirmed by comparing the relative wavelength positions and mode spacing under different pumping conditions. When the waveguide is half pumped, the mode spacing is doubled and the lasing wavelengths shift to the center of every two initial lasing modes. All these observations are consistent with the theoretical predictions and confirm the PT-symmetry breaking well.

**Introduction**

Soon after the pioneer work of Bender *et. al.* [1], parity-time (PT) symmetry has quickly gained tremendous attention due to its strong impacts in both fundamental researches and practical applications [2-12]. Fundamentally, it was shown that the eigenvalues of non-Hermitian Hamiltonians ($\widehat{H}^\dagger \neq \widehat{H}$) can be entirely real if they respect PT symmetry, $PT\widehat{H} = \widehat{H}PT$. Given a complex refractive index $n(x) = n_R(x) + i\, n_I(x)$, the necessary condition for a Hamiltonian to be PT symmetric is that $n_R(x) = n_R(-x)$ and $-n_I(x) = n_I(-x)$. In other words, the real parts of refractive index profile must be symmetric along x whereas the imaginary parts, i.e., gain and loss, should be anti-symmetric. Under these conditions, PT symmetry and PT symmetry breaking have been intensively studied in a number of optical systems, e.g., complex optical potentials [3], coupled waveguides [5], coupled cavities [10-13], and random media [11]. And many kinds of unique optical phenomena such as coherent perfect absorber [6, 9], loss induced transparency [7], unidirectional invisibility [8], and exceptional points in laser [14, 15] have been theoretically proposed and experimentally demonstrated.

In past few years, the concept of PT symmetry breaking has been further extended to optical microcavities in which gain and loss are equally applied in a manner that *n*(-*x*) = *n*\*(*x*) [10-12, 16-18]. Several types of unique properties have been discovered in such PT symmetric cavities, e.g., thresholdless PT symmetry breaking [16] and single-mode laser emission [17, 18]. The coexistence of PT symmetric laser and absorber is another prominent example [10-12]. In such systems, one mode is strongly amplified and the other mode is strongly attenuated simultaneously. Similar to the recently reported unconventional lasers [19, 20], the wavelengths of PT symmetric laser absorber are determined by the passive or lossy part of resonator and are not one-to-one correspondence with the resonances of passive cavity. In this sense, the mode numbers of PT symmetric laser absorber should be fewer and the mode spacing is doubled. Compared with the single-mode operation, this information can be an additional criterion to confirm the PT symmetry breaking. Till now, the researches of PT symmetric laser absorber are mostly limited in theoretical analysis and numerical calculations [10-12]. Very recently, the PT symmetric absorber has been experimentally observed in metamaterials at microwave frequency region [21]. The more interesting PT symmetric laser, however, has not been experimentally confirmed. In this article, we have studied the lasing actions in a dye doped stripe waveguide and experimentally confirmed the PT-symmetric laser absorber.

**Results and Discussions**

**PT symmetric laser absorber in stripe waveguide**

In conventional lasers with uniformly pumping, the lasing modes and resonances in passive cavities are one-to-one correspondence. Consequently, the imaginary parts of resonant frequencies change from negative values to positive values, indicating the transitions between leaky modes to lasing modes. This information can be clearly observed by finite element method based numerical simulations (Comsol Multiphysics 4.3a, the Radio Frequency module) [22]. In the numerical calculation, the waveguide is set as width $w = 1~\mu m$ and length $L = 10~\mu m$. The initial absorption is defined in the stripe waveguide by setting the imaginary parts of refractive index as -0.02 and the real part of refractive index of waveguide is 1.56. All these parameters are fixed in the following calculations. **Figure 1**(a) and 1(b) illustrate the behaviors of resonant frequency as a function of $n''$ (note the refractive index changes to $n(x) = n_R(x) + n_I(x) = n_R(x) + i \cdot n''$). Similar to the intuitive picture, the resonant frequencies of two resonances are almost independent to the gain, whereas their imaginary parts quickly transit from loss (Im(k) < 0) to gain (Im(k) > 0). Thus we know that the lasing modes under uniformly pumping are simply determined by the resonances within passive cavity [12, 23, 24]. This can be further confirmed by calculating the transmission spectra. As shown in **Figure 2**(a), while the transmitted intensity increases dramatically with the increase of gain, the peak positions and peak numbers are well consistent with resonances in passive cavity.

The case of PT symmetric laser is more attractive due to its unique characteristics. Following the necessary condition of PT symmetry, the gain and loss are equally distributed within the cavity in a manner of $n(-x) = n^*(x)$. Naively, one may think that the structure won't reach the laser threshold due to the balanced gain and loss. The recent studies have shown that the laser system can reach the PT symmetry breaking point, where a pair of resonance and antiresonance frequencies crosses the real frequency axis. This pair of resonances is known as PT symmetric laser absorber [6]. In general, the photon of PT symmetric laser mode is mostly localized within the gain region whereas the PT symmetric absorber mode is confined within the loss part. Figure 1(c) and 1(d) demonstrate their behaviors. We can see that the resonant frequencies of two nearby modes gradually approach to each other whereas their imaginary parts keep constant and bifurcate after the PT symmetry breaking. The corresponding transmission spectra under different $n''$ are plotted in Figure 2(b), where the mode number was halved and the mode spacing was doubled.

However, as shown in Refs. [17, 18], the perfect PT symmetry with equal gain and loss is hard to be experimentally fulfilled. For a practical experiment, the absorption in conventional laser system is usually fixed by the materials, whereas the amplification can be dramatically changed via applying different optical pumping or driven current. We then numerically studied the PT symmetric behaviors in such systems. In our calculation, the imaginary part of reflective index of lossy area (for 0 < x < 5 μm) is fixed at $n_I(x)$ = -0.02 and $n_I(x) = n''$ for gain area (-5 μm < x < 0). Figure 1(e) and (f) show one example of resonances within the quasi-PT symmetric stripe. With the increase of amplification ($n''$), two resonances gradually approach to each other. When $n'' > 0.065$, two resonances are very close in frequency. Meanwhile, the imaginary parts of frequencies increase first and then bifurcate at $n'' = 0.065$. The imaginary part of one mode decreases quickly and evolves to an absorption mode. In this sense, two resonances are still named as lasing mode and absorption mode. Their corresponding field patterns with $n'' = 0.07$ are shown in the inset of Figure 1(e). Similar to the conventional PT symmetric laser absorber, the lasing mode and absorption mode are mainly confined within the active and absorptive regions, respectively. In addition, the field distributions of two modes exponentially decay once their positions are away from the interface between gain and loss parts. This is also consistent with the recent reports of surface modes [12].

The difference between current study and conventional PT symmetric laser lies in the threshold and PT symmetric breaking point. In perfect PT symmetric system, PT symmetry breaking occurs around the point of threshold due to the balanced gain and loss. For the case of asymmetrical gain and loss, two resonances approach to the threshold (Im(k) = 0) with the increase of $n''$ from 0 to 0.0475 (see Figure 1(f)). However, the PT symmetry is broken at a larger $n''$. Consequently, while the lasing mode is continuously amplified by increasing the pumping (i.e., by increasing the $n''$), the absorption mode experiences two transitions between lossy mode and lasing mode. Therefore, instead of simply observing the lasing actions, the PT symmetry breaking should be validated with more caution. Due to the correspondence between lasing modes and transmission spectra with gain and loss, we have also studied the light transmission through the stripe waveguide. All the results are summarized in Figure 2(c). We can see that all the modes gradually reach the threshold and lase as the increase of $n''$. Soon after the laser threshold, the intensity of one mode keeps increasing whereas the intensity of the other one reduces gradually and finally vanishes due to the strong absorption. Consequently, such kind of practical system has separated threshold point and bifurcating point. And the mode spacing can finally be doubled at higher pumping power (larger $n''$ in simulation). Therefore, the threshold point, bifurcating point, and mode spacing are the key parameters to determine the PT symmetry breaking. In addition, the separation distance between threshold and bifurcating points is inversely dependent on the fixed absorption in the leaky part. Thus the deviation from perfect PT symmetry in real experiment can also be roughly estimated.

**Experimental observation of PT symmetric laser absorber**

Based on above theoretical analysis and numerical calculation, we thus fabricated stripe waveguide to experimentally observe the PT symmetric laser. The stripe waveguides in the present study were fabricated with standard photolithography on commercially available GM1050 doped with 0.5 wt% Rhodamine B [25]. The photoresist was spin-coated onto substrate with 4000 r/min for 40 s. The measured thickness of photoresist was around 2 microns. The substrate was commercial glass sheet with refractive index ~ 1.5 to ensure the light confinement in vertical direction. The exposure density and exposure time were 5 mW/cm$^2$ and 4 min. After 22 s developments in PGMEA, dye doped stripe waveguides were obtained. The microscope image of fabricated waveguide is shown as inset in **Figure 3**(d). From the inset, we can see that the end of waveguide shows rectangular shape and the sidewalls are quite smooth. The width of waveguide is around 5 μm and its total length is about 1mm.

The waveguides were mounted on a three-dimensional translation stage and pumped by a frequency doubled Nd:YAG laser (532nm, 7ns pulse duration). The laser spot was focused by a cylindrical lens to 0.2 mm width laser stripe. The direction of laser stripe followed the waveguide well. During the experiment, the laser stripe and waveguides were fixed and the pumping lengths were controlled by an variable slit. The emitted lasers were collected by a lens and then coupled into a spectrometer and a CCD camera (Princeton instrument) through a multimode fiber. A long pass filter was placed before the fiber to exclude the influences of pump lasers. In our experiment, entirely pumping and half pumping configurations have been utilized, as shown in Figure 3(a) and 3(c). When the stripe waveguide was entirely pumped at 23 μJ, the laser spectra were recorded and shown in Figure 3(b), where periodic modes could be observed. The mode spacing (Δλ) is 0.112 nm, which matches the Fabry-Perot cavity modes (Δλ~$\frac{\lambda^2}{2nL} = 0.117\ nm$) well. Once the stripe waveguides was half pumped, the emission spectrum was totally different. One example is depicted in Figure 3(d). When the left part of stripe waveguide was pumped by a 0.5 mm laser beam (see schematic picture in Figure 3(c)), periodic laser peaks can still be observed in the laser spectrum. The differences between Figure 3(b) and 3(d) are the wavelengths of laser peaks and their mode spacing. It is easy to see that peak-a under half pumping is right in the middle of nearby modes 1 and mode 2 under entirely pumping. Similar wavelength shifts hold true for the other peaks –b and –c in Figure 3(d). Most interestingly, the mode number in Figure 3(d) is also half of that in Figure 3(b), giving a doubled mode spacing. All these phenomena are consistent with the numerical calculations in Figure 1 well and thus indicate the PT symmetry breaking well.

In experiment, the optical pumping or current injection can also change the local refractive index of microcavity. To exclude the influences of optical pumping, we further studied the dependences of laser spectra on the pumping power. By fixing the pumping length to half of the stripe waveguide, we measured the laser spectra under different pumping powers. When the pumping energy was below 10 μJ, the laser spectrum was dominated by spontaneous emission, which was almost on the noise level in the high resolution spectrum. Once the pumping energy was above 10 μJ, periodic laser peaks emerged in the spectrum (see **Figure 4**(a)) and the total intensity increased dramatically. From the dependence of output on pumping power, a clear laser threshold can be observed around 10 μJ.

The mode spacing of the laser spectrum around threshold was 0.112 nm. From the mode spacing, we knew that the lasers around threshold were conventional Fabry-Perot lasers within the cavities. Such conventional lasing modes with $\Delta\lambda \sim 0.112$ nm were well kept in a wide range of pumping power, only with some intensity changes. As shown in Figure 4(a), the relative intensities of every two modes kept decreasing. The summary of energy-dependent intensities was plotted in Figure 4(d) as a function of pumping energy. We can see that two nearby modes have similar intensities around the threshold. With the increase of pumping power, the intensity of one mode increased quickly, whereas the other mode decreased and finally disappeared at pumping power P > 20 μJ. This behavior is quite similar to the numerical calculations in Figure 1(b), where two resonances bifurcated after the threshold. This bifurcating behavior can easily exclude the possibility of the pumping induced change in real part of refractive index. Another important information in Figure 4(d) is that slight separation between threshold point and bifurcating point is very small, indicating that the laser in our experiments is very close to perfect PT symmetric lasers. Meanwhile, the lasing wavelengths were also found to be dependent on pumping power. As shown in Figure 4(c), while all the resonances shifted with the pumping power, the separation distances between every two modes slightly changed and finally turned to be one mode in the middle of initial two modes. This is also consistent with the theoretical predictions.

**Conclusion**

In summary, we have fabricated dye doped stripe waveguides and studied their lasing actions under different pumping conditions. We found that the lasing wavelengths shifted and mode spacing were doubled when the pumping conditions of waveguides were changed from entirely pumping to half pumping. With the increase of pumping power under half pumping, we also observed the bifurcation in intensities between two nearby lasing modes. From the slight difference between threshold point and bifurcating point, the laser emissions in stripe waveguide were considered as PT symmetric lasers. All these observations matched the theoretical predictions and numerical calculations well. We believe these results will be interesting for the applications of unconventional lasers.

**References:**

[1]     C. M. Bender, S. Boettcher, *Phys. Rev. Lett.* **80**, 5243-5246 (1998).

[2]     R. El-Ganainy, K. G. Makris, D. N. Christodoulides, Z. H. Musslimani, *Opt. Lett.* **32**, 2632 (2007).

[3]     K. G. Makris, R. El-Ganainy, D. N. Christodoulides, Z. H. Musslimani, *Phys. Rev. Lett.* **100**, 103904 (2008).

[4]     C. E. Rütter, K. G. Makris, R. El-Ganainy, D. N. Christodoulides, M. Segev, D. Kip, *Nature Phys.* **24**, 192-195 (2010).

[5]     A. Ruschhaupt, F. Delgado, J. G. Muga, *J. Phys. A* **38**, L171-176 (2005).

[6]     Y. D. Chong, L. Ge, H. Cao, A. D. Stone, *Phys. Rev. Lett.* **105**, 053901 (2010).

[7]     L. Feng, Y. -L. Xu, W. S. Fegadolli, M-H. Lu, J. E. B. Oliveira, V. R. Almeida, Y-F. Chen, A. Scherer, *Nature Mat.* **12**, 108-113 (2013).

[8]     B. Peng, S. K. Özdemir, F. C. Lei, F. Monifi, M. Gianfreda, G. L. Long, S. H. Fan, F. Nori, C. M. Bender, L. Yang, *Nature Phys.* **10**, 394-398 (2014).

[9]     W. Wan, Y. D. Chong, L. Ge, H. Noh, A. D. Stone, H. Cao, *Science* **331**, 889 (2011).

[10]    S. Longhi, *Phys. Rev. A* **82**, 031801® (2009).

[11]  Y. D. Chong, L. Ge, A. D. Stone, *Phys. Rev. Lett.* **106**, 093902 (2011).

[12]  L. Ge, Y. D. Chong, S. Rotter, H. E. Türeci, A. D. Stone, *Phys. Rev. A* **84**, 023820 (2011).

[13]  B. Peng, S. K. Özdemir, S. Rotter, H. Yilmaz, M. Liertzer, F. Monifi, C. M. Bender, F. Nori, L. Yang, *Science* **346**, 328-332 (2014).

[14]  M. Liertzer, L. Ge, A. Cerjan, A. D. Stone, H. E. Türeci, S. Rotter, *Phys. Rev. Lett.* **108**, 173901 (2012).

[15]  M. Brandstetter, M. Liertzer, C. Deutsch, P. Klang, J. Schöberl, H. E. Türeci, G. Strasser, K. Unterrainer, S. Rotter, *Nature Commun.* **5**, 4034 (2014).

[16]  L. Ge, *Phys. Rev. X* **3**, 031011 (2014).

[17]  L. Feng, Z. J. Wong, R. -M. Ma, Y. Wang, X. Zhang, *Science* **346**, 972-975 (2014).

[18]  H. Hodaei, M. Miri, M. Heinrich, D. N. Christodoulides, M. Khajavikan, *Science* **346**, 975-978 (2014).

[19]  J. Andreasen, H. Cao, *Opt. Lett.* **34**, 3586-3588 (2009).

[20]  J. Andreasen, C. Vanneste, L. Ge, H. Cao, *Phys. Rev. A* **81**, 043818 (2010).

[21]  Y. Sun, W. Tan, H. Q. Li, J. Li, H. Chen, *Phys. Rev. Lett.* **112**, 143903 (2014).

[22]  Z. Y. Gu, S. Liu, S. Sun, K. Y. Wang, Q. Lyu, S. M. Xiao, Q. H. Song, *Sci. Rep.* **5**, 9171 (2015).

[23] X. F. Jiang, Y. F. Xiao, C. L. Zou, L. N. He, C. H. Dong, B. B. Li, Y. Li, F. W. Sun, L. Yang, Q. H. Gong, *Adv. Mater.* **24**, OP260 (2012).

[24] Q. H. Song, L. Ge, B. Redding, H. Cao, *Phys. Rev. Lett.* **108**, 243902 (2012).

[25] M. Li, N. Zhang, K. Y. Wang, J. K. Li, S. M. Xiao, Q. H. Song, arXiv: 1504.03436 (2015).

**Acknowledgements**

This work is supported by NSFC11204055, NSFC61222507, NSFC11374078, NCET-11-0809, Shenzhen Peacock plan under the Nos. KQCX2012080709143322 and KQCX20130627094615410, and Shenzhen Fundamental research projects under the Nos. JCYJ20130329155148184, JCYJ20140417172417110, JCYJ20140417172417096.

**Key words:** Parity-time symmetry, stripe waveguides, laser and absorber

**Figure Caption:**

**Fig.1**: The dependences of real and imaginary parts of conventional lasers (a, b), PT symmetric lasers (c, d), and loss fixed systems (e, f). Here the initial loss is zero in (a) - (d) and $n_I(x) = -0.02$ for $0 < x < 5$ μm in (e) and (f). The width and length of stripe waveguide are *1 μm* and *10 μm*, respectively. The top and bottom panels in the insets of (e) are the field patterns of lasing mode and absorption mode with $n'' = 0.07$, respectively.

**Fig.2**: The transmission spectra in conventional system (a), PT symmetric system (b), and loss fixed system (c) under different $n''$. The transmission peak positions are fixed in conventional laser system. The mode number reduces and mode spacing is doubled when PT symmetric system lase. In loss fixed system, all the resonances first lase and then one mode disappear to reduce the mode number and double the mode spacing.

**Fig.3**: Laser spectrum under entirely pumping and half pumping. (a) and (c) are the schematic pictures of different pumping configurations. (b) and (d) are their corresponding laser spectra. Here the pumping power was 23 μJ. The length of laser stripe was 1 mm. The inset in (d) is the microscope image of the sample. The width and length of waveguide are 5 μm and 1mm, respectively.

**Fig.4**: Laser behaviors as a function of pump power under half pumping. (a) The laser spectra under different pumping power, (b) the threshold behaviors, (c) and (d) are the dependences of lasing wavelengths and intensities on pumping power. The modes $A_1$, $B_1$, $A_2$, and $B_2$ in (d) are marked in (c).

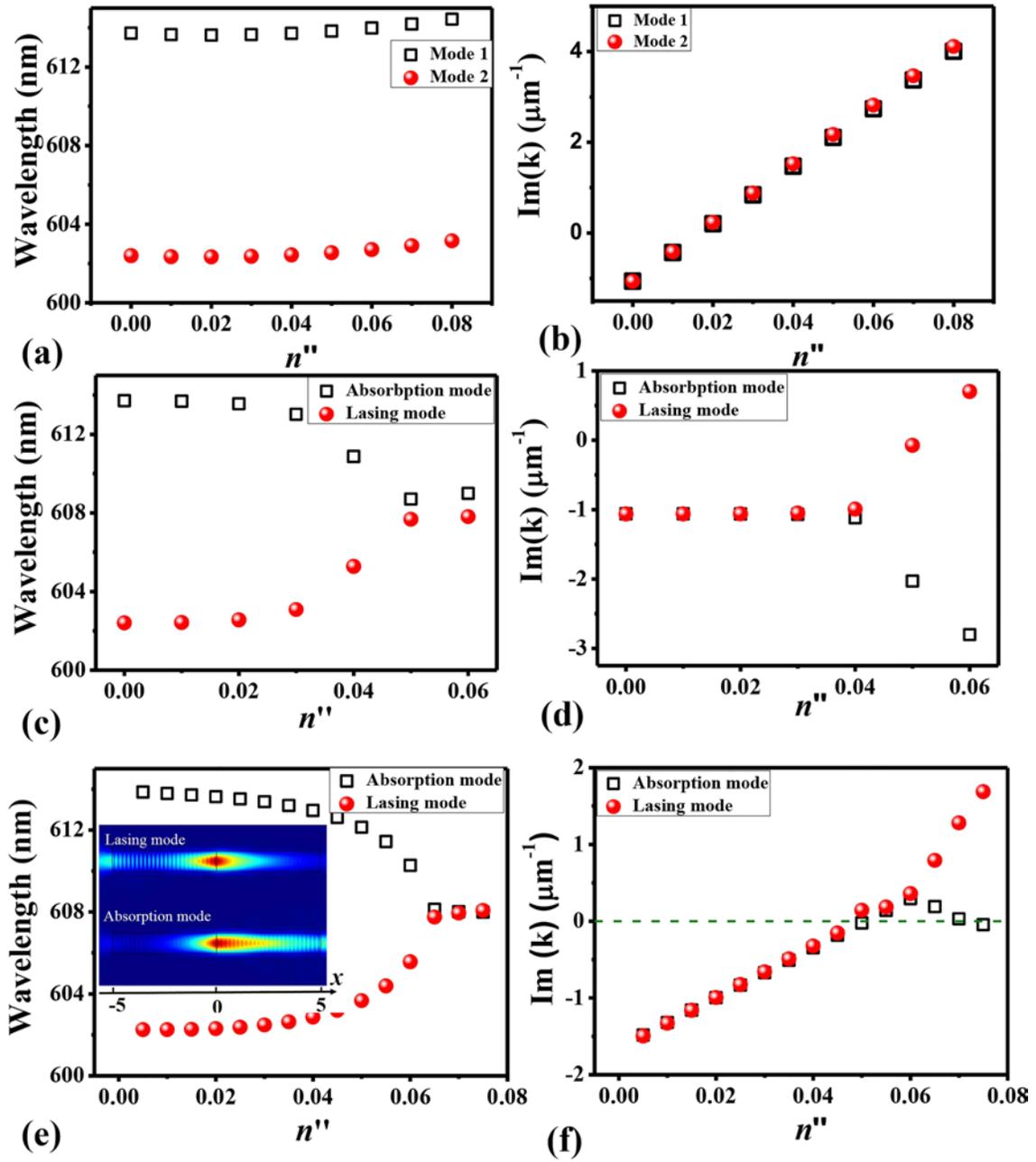

Fig. 1

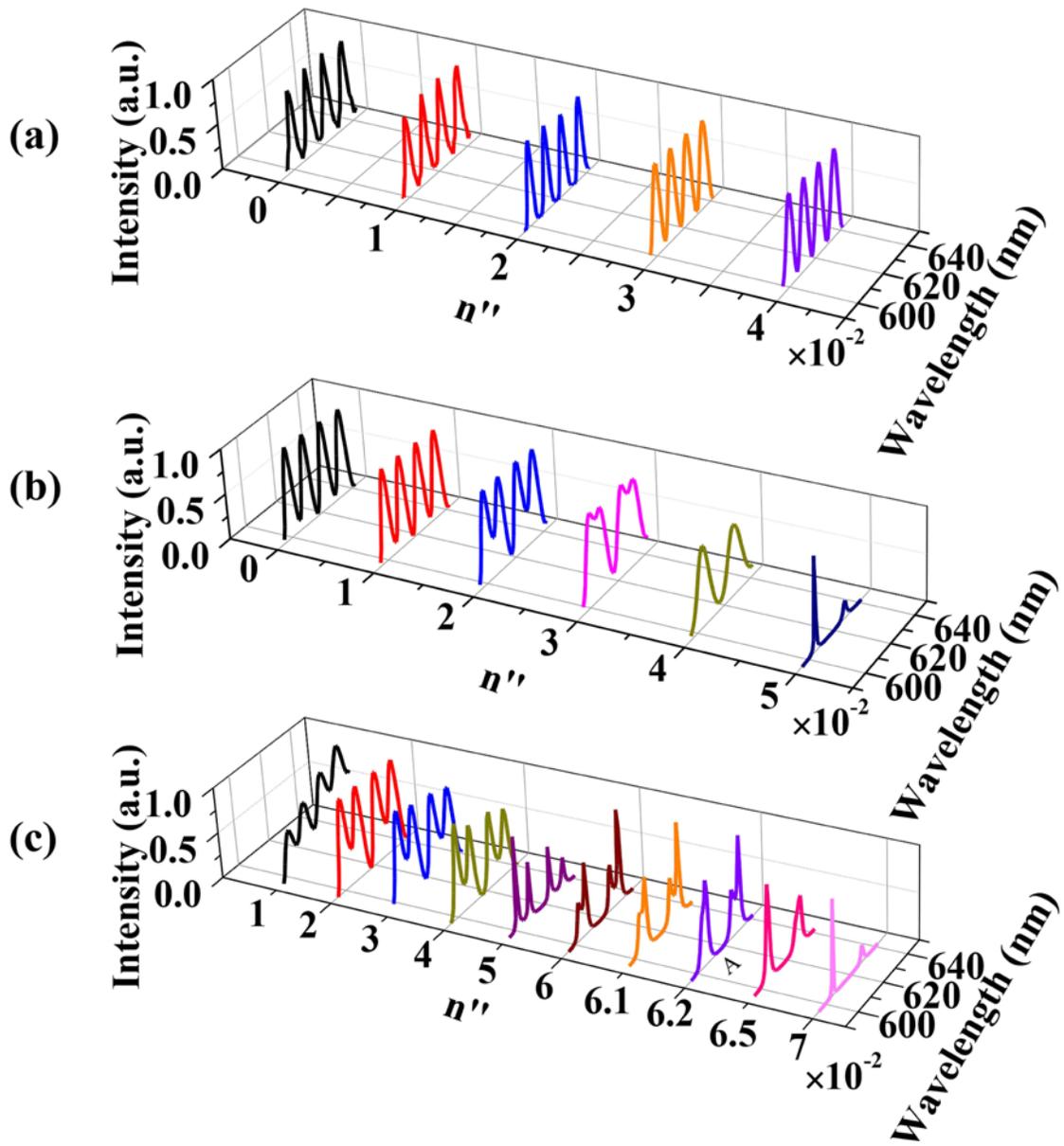

Fig. 2

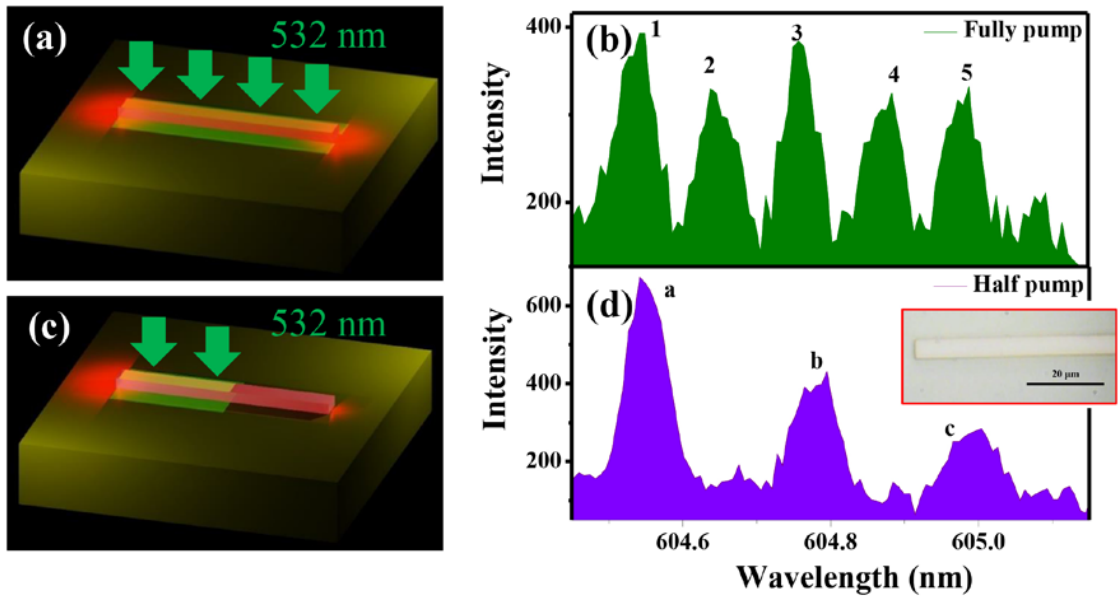

Fig. 3

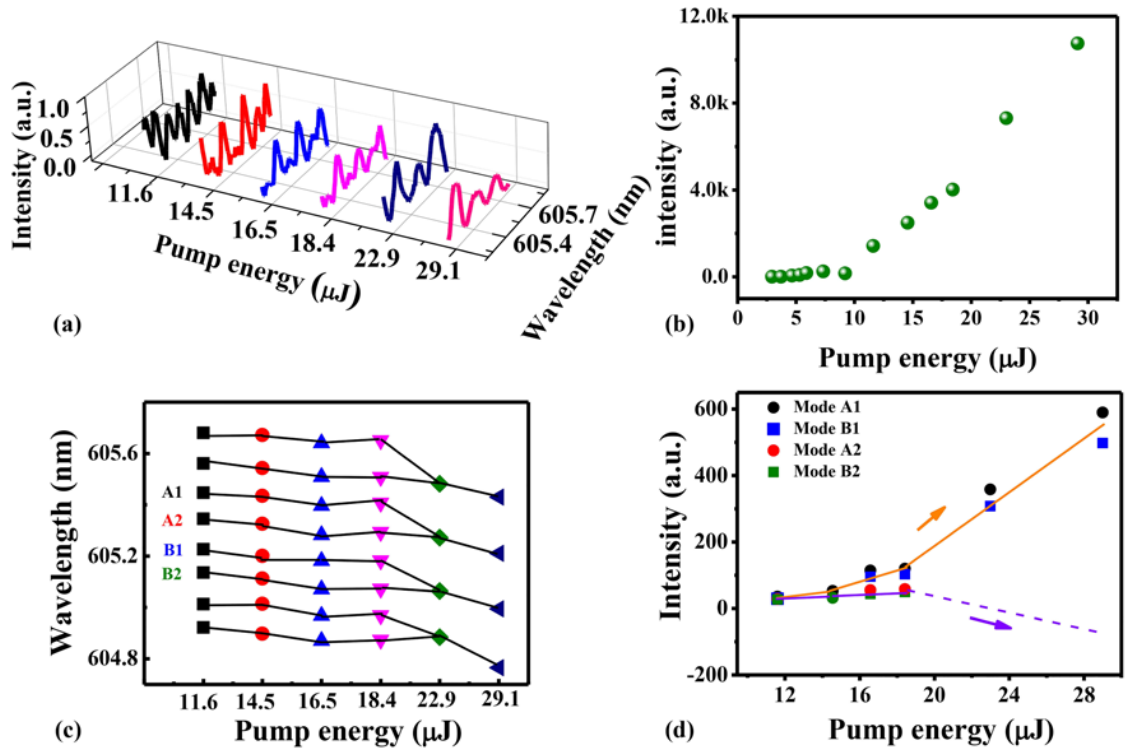

Fig.4